\documentclass{iau} 
\usepackage{graphicx}

\title[Optical Interferometry and Gaia on Cepheids] 
{Optical interferometry and Gaia parallaxes for a robust calibration of the Cepheid distance scale}

\author[Pierre Kervella \etal\ ]   
{Pierre Kervella$^{1, 2}$,
 Antoine M\'erand$^3$,
Alexandre Gallenne$^4$,
Boris Trahin$^{1}$,
Simon Borgniet$^{1}$,
Grzegorz Pietrzynski$^{5, 7}$,
Nicolas Nardetto$^{6}$
\and Wolfgang Gieren$^{7}$
 }

\affiliation{$^1$Unidad Mixta Internacional Franco-Chilena de Astronom\'{i}a (CNRS UMI 3386), Departamento de Astronom\'{i}a, Universidad de Chile, Santiago, Chile \\[\affilskip]
$^2$LESIA (UMR 8109), Observatoire de Paris, PSL Research University, CNRS, UPMC, Univ. Paris-Diderot, 92195 Meudon, France, email: {\tt pierre.kervella@obspm.fr} \\[\affilskip]
$^3$European Southern Observatory, Karl-Schwarzschild-str. 2, D-85748 Garching, Germany.\\[\affilskip]
$^4$European Southern Observatory, Alonso de C\'ordova 3107, Casilla 19001, Santiago 19, Chile.\\[\affilskip]
$^5$ Nicolaus Copernicus Astronomical Center, Polish Academy of Sciences, Warszawa, Poland\\[\affilskip]
$^6$ Laboratoire Lagrange, UMR7293, Universit\'e de Nice Sophia-Antipolis, CNRS, Observatoire de la C\^ote d'Azur, 06000 Nice,
France\\[\affilskip]
$^7$ Universidad de Concepci\'on, Departamento de Astronom\'ia, Casilla 160-C, Concepci\'on, Chile
}

\pubyear{2017}
\volume{330}  
\jname{Astrometry and Astrophysics in the Gaia sky}
\editors{A. Recio-Blanco, P. de Laverny, A. Brown, \& T. Prusti, eds.}
\begin{document}

\maketitle

\begin{abstract}
We present the modeling tool we developed to incorporate multi-technique observations of Cepheids in a single pulsation model: the Spectro-Photo-Interferometry of Pulsating Stars (SPIPS). The combination of angular diameters from optical interferometry, radial velocities and photometry with the coming Gaia DR2 parallaxes of nearby Galactic Cepheids will soon enable us to calibrate the projection factor of the classical Parallax-of-Pulsation method. This will extend its applicability to Cepheids too distant for accurate Gaia parallax measurements, and allow us to precisely calibrate the Leavitt law's zero point. As an example application, we present the SPIPS model of the long-period Cepheid RS Pup that provides a measurement of its projection factor, using the independent distance estimated from its light echoes.

\end{abstract}

\firstsection 
\section{Introduction}

One century after the discovery of their Period-Luminosity relation (the Leavitt law) by \cite[Leavitt \& Pickering~(1912)]{Leavitt12}, Cepheids are still the keystone of the empirical cosmic distance ladder.
However, due to the large distances of Cepheids (particularly the long-period pulsators), only relatively imprecise parallax measurements are available (e.g., from Hipparcos; \cite[van Leeuwen \etal\ ~2007]{vanLeeuwen07}).
As a result, the calibration of the zero point of the Leavitt law using Galactic Cepheids is insufficient accurate.
Gaia's DR2 high accuracy parallaxes will bring a tremendous improvement, with distances to hundreds of individual Cepheids measured to a few percent accuracy. 
Alternatively, a classical avenue to measure the distances to individual Galactic and LMC Cepheids is the Parallax-of-Pulsation (PoP) method, also known as the Baade-Wesselink (BW) technique.
We here briefly present the SPIPS modeling tool that we developed to reproduce the classical observables of the pulsation of Cepheids (radial velocity, photometry, interferometry,...), and how we plan to use it in conjunction with the Gaia parallaxes to calibrate the PoP technique and eventually the Leavitt law.

\section{The Parallax-of-Pulsation technique and SPIPS \label{SPIPS}}

\begin{figure}[t]
\begin{center}
 \includegraphics[width=12cm]{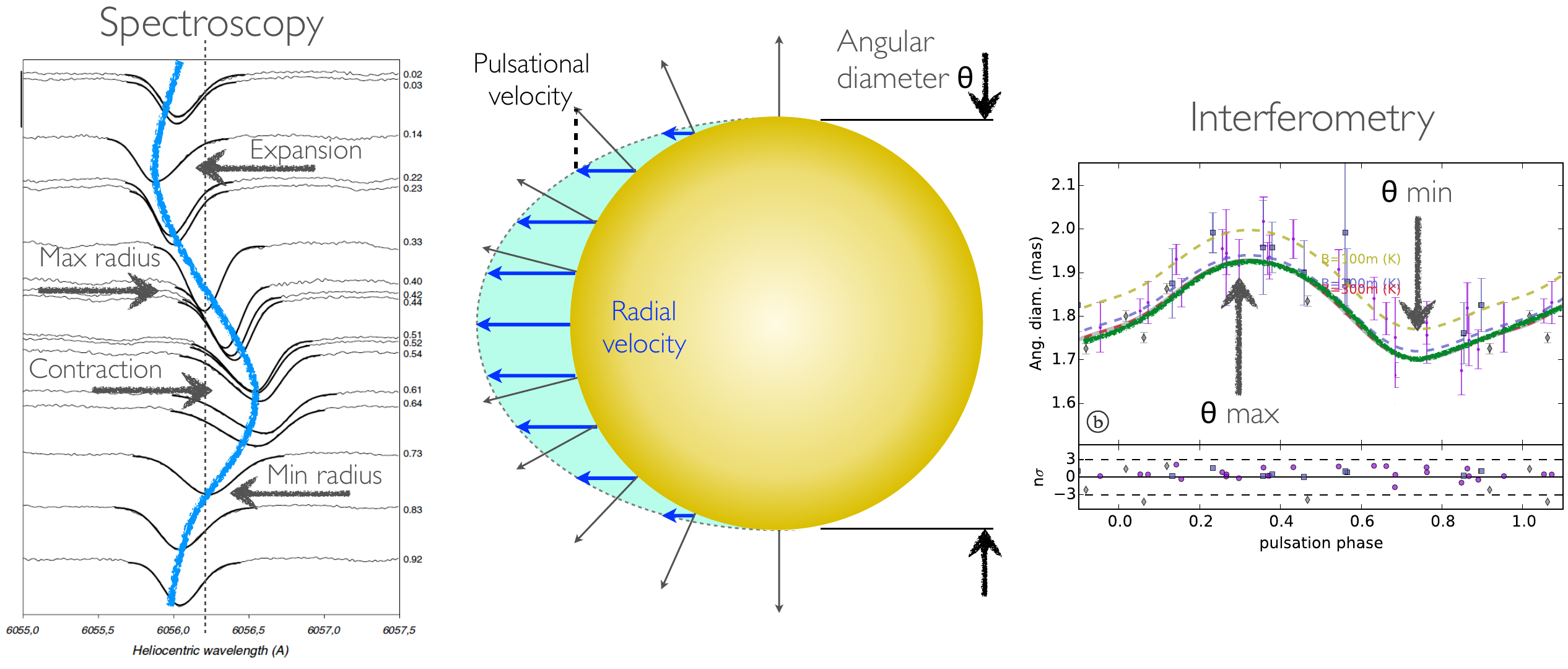} 
 \caption{Principle of the classical Parallax-of-Pulsation distance determination on the Cepheid $\beta$\,Dor.
 {\it Left panel:} Doppler displacement of the Fe I line at {6056\,\AA} obtained from high resolution HARPS spectra (\cite[Nardetto \etal\ 2006]{Nardetto06}). 
 {\it Right panel:} Interferometric angular diameter as a function of the pulsation phase (\cite[Breitfelder \etal\ 2016]{Breitfelder16}).
 }
   \label{fig1}
\end{center}
\end{figure}

The PoP technique relies on the comparison of
(1) the amplitude of the pulsation of the star from the integration of its radial velocity curve measured using spectroscopy and
(2) the change in angular diameter estimated from its brightness and color, or measured using interferometry.
The distance of the Cepheid is then obtained by fitting the linear and angular amplitudes (see, e.g.,~\cite[Storm \etal\ 2011]{Storm11}).
Although potentially precise at the percent level, the major weakness of the PoP technique is that it uses a numerical factor to convert disk-integrated radial velocities into photospheric velocities: the projection factor ($p$-factor; \cite[Nardetto \etal\ 2017]{2017Nardetto}).
It is defined as the ratio of the pulsational velocity over the radial velocity projected on the line of sight.
The absolute accuracy of the PoP technique is currently limited by our knowledge of the $p$-factor, that is degenerate with the distance $d$ (the derived quantity is $d/p$). This parameter is particularly complex to model, as it represents simultaneously the spherical geometry, the limb darkening, and the gas dynamics in the Cepheid atmosphere.

The \textit{Spectro Photo Interferometry of Pulsating Stars} (SPIPS; \cite[M\'erand \etal\ 2015]{Merand15}) model of a Cepheid is built assuming that the star is a radially pulsating sphere, for which the pulsational velocity $v_\mathrm{puls}(t)$ and the effective temperature $T_\mathrm{eff}(t)$ are the basic parameters.
The cyclic variation of these parameters is represented using classical Fourier series or periodic splines functions.
The photometry in all filters is computed using ATLAS9 atmosphere models, considering their bandpasses and zero points.
The interstellar reddening is parametrized using the standard $E(B-V)$ color excess for Milky Way dust ($R_V=3.1$) and computed for the phase-variable $T_\mathrm{eff}$.
Angular diameters including limb darkening are produced by the model to match the interferometric measurements, using SATLAS models (\cite[Neilson \etal\ 2013]{2013Neilson}).
Finally, circumstellar envelopes are included in the modeling (\cite[Gallenne \etal\ 2012]{2012Gallenne}; \cite[Gallenne \etal\ 2013]{2013Gallenne}), as well as the presence of stellar companions (\cite[Gallenne \etal\ 2015]{2015Gallenne}; \cite[Gallenne \etal\ 2017]{2017Gallenne}).
SPIPS can be employed with only photometry and radial velocity, but interferometric angular diameters are independent of interstellar absorption, allowing to de-correlate the reddening and the effective temperature.
Angular diameters thus bring a considerable added value to the quality and robustness of the derived parameters.
Recent applications of SPIPS to Cepheids can be found in \cite[M\'erand \etal\  (2015)]{Merand15} ($\delta$\,Cep, $\eta$\,Aql), \cite[Breitfelder \etal\ (2016)]{Breitfelder16} (Cepheids with HST/FGS parallaxes) and \cite[Breitfelder \etal\ (2015)]{Breitfelder15} (Type 2 Cepheid $\kappa$\,Pav).

\section{The projection factor of RS\,Puppis and the $p-P$ relation}

\begin{figure}[t]
\begin{center}
 \includegraphics[width=13cm]{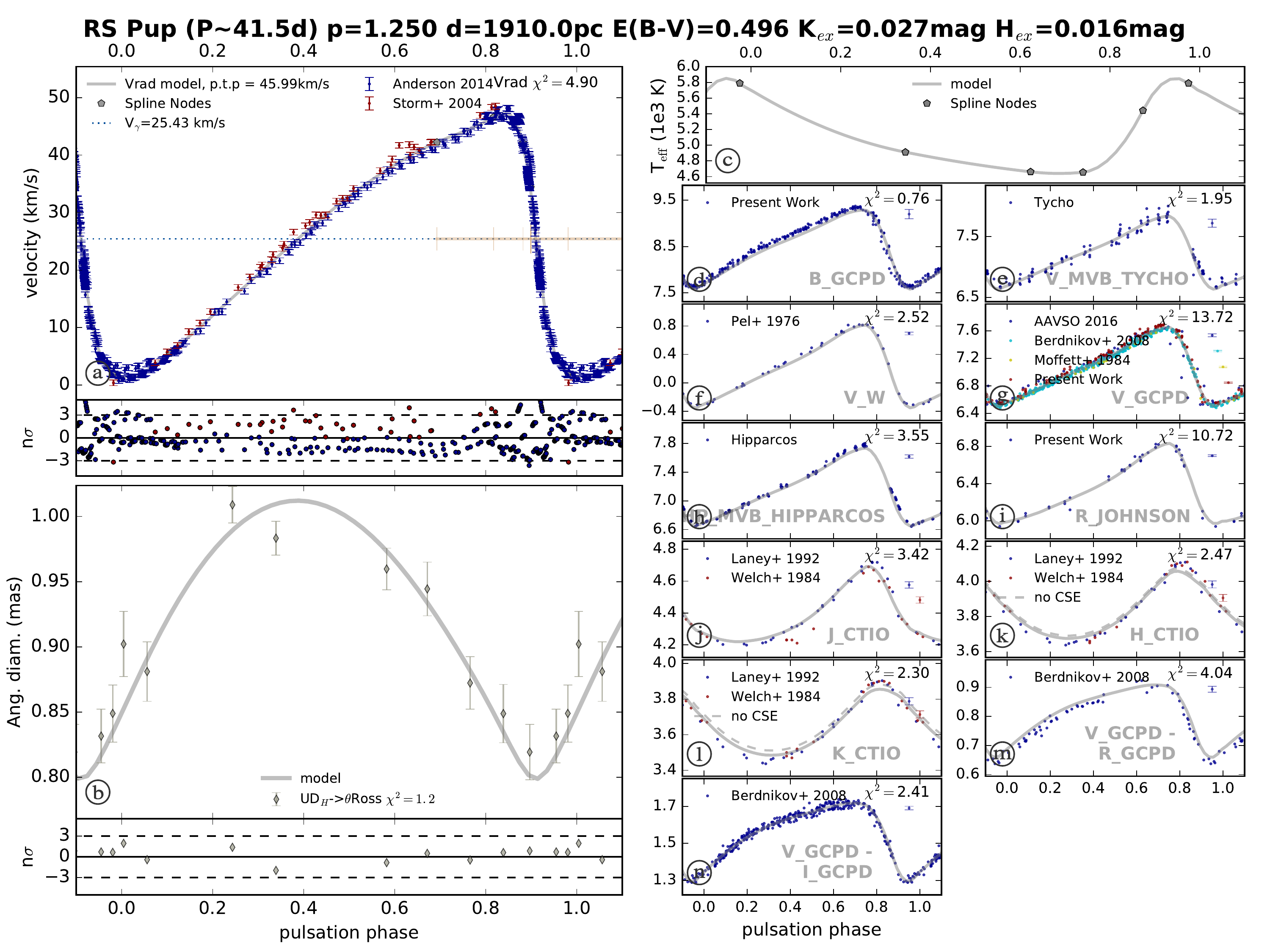} 
 \caption{SPIPS combined fit of the observations of RS Puppis ($P = 41.5$\,days).
The radial velocity measurements are presented in the upper left panel, and the interferometric angular diameters in the lower left panel. The model's effective temperature is shown in the upper right panel, and the photometry in the different sub-panels of the right column (\cite[Kervella \etal\ 2017]{Kervella17}).}
   \label{fig2}
\end{center}
\end{figure}

\begin{figure}[h]
\begin{center}
 \includegraphics[width=7.5cm]{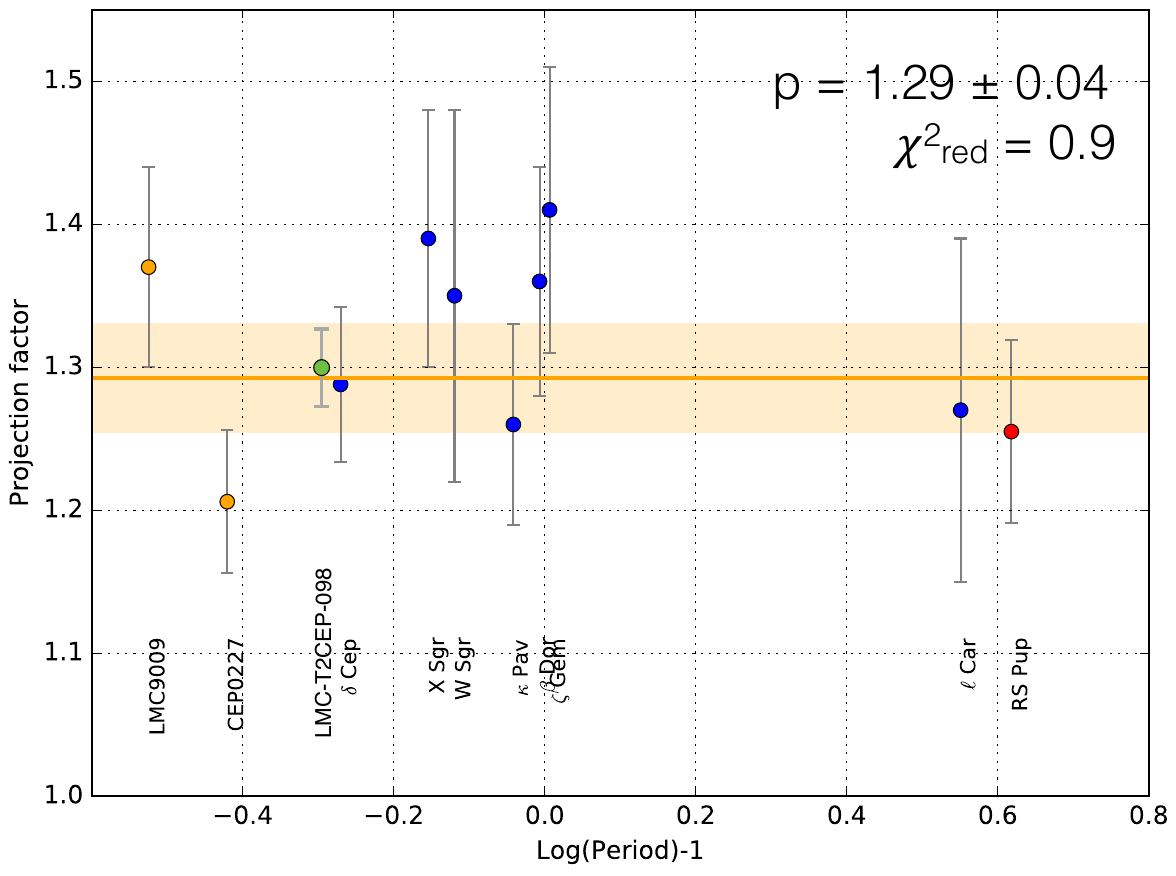} 
 \caption{Measured p-factors of Cepheids (from \cite[Kervella \etal\ 2017]{Kervella17}). The green point is the measurement of the binary type 2 Cepheid OGLE-LMC-T2CEP-098 by \cite[Pilecki \etal\ (2017)]{Pilecki17}.
}
   \label{fig3}
\end{center}
\end{figure}

RS\,Pup is one of the intrinsically brightest Cepheids in the Galaxy, and it is remarkable due to its large circumstellar nebula that reflects the light variations of the Cepheid.
From the light echoes, \cite[Kervella \etal\ (2014)]{2014Kervella} derived its distance of $d = 1910 \pm 80$\,pc.
Knowing the distance resolves the intrinsic distance/$p$-factor degeneracy of SPIPS and gives access to $p$.
The pulsation model (Fig.~\ref{fig2}) has been presented by \cite[Kervella \etal\ (2017)]{2017Kervella}, giving a best-fit $p$-factor of $p = 1.25 \pm 0.06$ ($\pm 5\%$).
The addition of the $p$-factor of RS\,Pup to the limited set of existing $p$-factor measurements is particularly valuable, as it was up to now mostly limited to short and intermediate period Cepheids (Fig.~\ref{fig3}).
The other exception is $\ell$\,Car ($P=35.5$\,days), whose parallax was measured by \cite[Benedict \etal\ (2007)]{2007Benedict}.
The $p$-factors of $\ell$\,Car and RS\,Pup are statistically identical within their uncertainties.
The simple model of a constant $p = 1.29 \pm 0.04$ over the sampled range of Cepheid periods (3.0 to 41.5\,days) is consistent with the observations ($\chi^2_\mathrm{red}=0.9$; \cite[Kervella \etal\ 2017]{2017Kervella}).

\section{The contribution of Gaia}

Gaia's DR2 in April 2018 will considerably improve the accuracy of Galactic Cepheid parallaxes compared to DR1 (\cite[Casertano \etal\  2017]{2017Casertano}; \cite[Gaia Collaboration 2017]{Gaia2017}).
Based on the foreseen accuracy, more than 400 Cepheid parallaxes will be measured with an accuracy $<3\%$, out of which approximately one hundred to $<1\%$.
The observations of a sample of 18 long-period Cepheid using the spatial scanning mode of the HST/WFC3 by \cite[Casertano \etal\  (2016)]{2016Casertano} will complement this data set.
This unprecedented catalog will make it possible to obtain an extremely accurate calibration of the Leavitt law and other fundamental properties of Cepheids.
However, to reach this accuracy implies that the pulsation models employed to interpret the observations must also reach this level of accuracy.
The flexibility of the SPIPS modeling enables the integration of all types of new and archival observations in a consistent and robust approach. This will contribute to mitigate the influence of various systematics of observational origin, and bring the accuracy of the calibration of the Cepheid distance scale down to $1\%$.

\noindent \textit{Acknowledgements:}
{\small The authors acknowledge the support of the French Agence Nationale de la Recherche (ANR-15-CE31-0012-01 UnlockCepheids).
W.G. and G.P. gratefully acknowledge the BASAL Centro de Astrofisica y Tecnologias Afines (CATA) PFB-06/2007. 
W.G. acknowledges support from the Chilean Millenium Institute of Astrophysics (MAS) IC120009.
This research has received funding from the European Research Council (Horizon 2020 grant No 695099).}

%


\begin{thebibliography}{}



\bibitem[Benedict \etal\ (2007)]{2007Benedict} {Benedict, G. F., McArthur, B. E., Feast, M. W., \etal} 2007, \textit{AJ}, 133, 1810

\bibitem[Breitfelder \etal\ (2015)]{Breitfelder15}{Breitfelder, J., Kervella, P., M\'erand, A., \etal.} 2015 \textit{A\&A}, 576, A64

\bibitem[Breitfelder \etal\ (2016)]{Breitfelder16} {Breitfelder, J., M\'erand, A., Kervella, P., \etal} 2016, \textit{A\&A}, 587, A117 

\bibitem[Casertano \etal\ (2016)]{2016Casertano} {Casertano, S., Riess, A. G., Anderson, J., \etal} 2016, \textit{ApJ}, 825, 11

\bibitem[Casertano \etal\ (2017)]{2017Casertano} {Casertano, S., Riess, A. G., Bucciarelli, B., \& Lattanzi, M. G.} 2017, \textit{A\&A}, 599, A67

\bibitem[Gaia Collaboration (2017)]{Gaia2017}{Gaia~Collaboration, Clementini, G., Eyer, L., \etal} 2017, \textit{A\&A}, in press, arXiv:1705.00688

\bibitem[Gallenne \etal\ (2012)]{2012Gallenne} {Gallenne, A., M\'erand, A., Kervella, P.} 2012, \textit{A\&A}, 538, A24

\bibitem[Gallenne \etal\ (2013)]{2013Gallenne} {Gallenne, A., M\'erand, A., Kervella, P., \etal\ } 2013, \textit{A\&A}, 558, A140

\bibitem[Gallenne \etal\ (2015)]{2015Gallenne} {Gallenne, A., M\'erand, A., Kervella, P., \etal\ } 2015, \textit{A\&A}, 579, A68

\bibitem[Gallenne \etal\ (2017)]{2017Gallenne} {Gallenne, A., Kervella, P., M\'erand, A., Evans, N. R., \& Proffitt, C.} 2017, these proceedings


\bibitem[Kervella \etal\ (2014)]{2014Kervella}{Kervella, P., Bond, H. E., Cracraft, M., \etal} 2014, \textit{A\&A}, 572, A7

\bibitem[Kervella \etal\ (2017)]{Kervella17} {Kervella, P., Trahin, B., Bond, H. E., \etal} 2017, \textit{A\&A}, 600, A127


\bibitem[Leavitt \& Pickering (1912)]{Leavitt12}{Leavitt, H. S. \& Pickering, E. C.} 1912, \textit{Harvard College Observatory Circular},
173, 1

\bibitem[M\'erand \etal\ (2015)]{Merand15} {M\'erand, A., Kervella, P., Breitfelder, J., \etal} 2015, \textit{A\&A}, 584, A80

\bibitem[Nardetto \etal\ (2004)]{Nardetto06} {Nardetto, N., Mourard, D., Kervella, P., \etal} 2006, \textit{A\&A}, 453, 309 


\bibitem[Nardetto \etal\ (2017)]{2017Nardetto} {Nardetto, N., Poretti, E., Rainer, M., \etal} 2017, \textit{A\&A}, 597, A73

\bibitem[Neilson \etal\ (2013)]{2013Neilson}{Neilson, H. R., \& Lester, J. B.} 2013, \textit{A\&A}, 554, A98


\bibitem[Pilecki \etal\ (2017)]{Pilecki17} {Pilecki, B., Gieren, W., Smolec, R., \etal} 2017, \textit{ApJ}, in press (arXiv:1704.07782)

\bibitem[Storm \etal\ (2011)]{Storm11} {Storm, J., Gieren, W., Fouqu\'e, P., \etal} 2011, \textit{A\&A}, 534, A94



\end{thebibliography}
\end{document}